\documentclass[useAMS,usenatbib,usegraphicx]{mn2e}

\usepackage{float}
\usepackage{color}
\usepackage{graphicx}
\usepackage{natbib}
\usepackage{lscape}
\usepackage{amssymb}

\title[H{\small I}\ towards Puppis A]{A high resolution H{\sc i} study 
towards the supernova remnant Puppis A and its environments}

\author[Reynoso, Cichowolski, and Walsh]{E. M. Reynoso$^{1}$\thanks{Email:
ereynoso@iafe.uba.ar}, S. Cichowolski$^{1}$\thanks{Email: scicho@iafe.uba.ar}, 
	and
        A. J. Walsh$^{2}$\thanks{Email: andrew.walsh@curtin.edu.au}\\
        $^{1}$Instituto de Astronom\'\i a y F\'\i sica del Espacio (IAFE), 
	UBA-CONICET, Av. Int. G\"uiraldes 2620, Pabell\'on IAFE, Ciudad 
	Universitaria, \\Ciudad Aut\'onoma de Buenos Aires, Argentina\\ 
	$^{2}$International Centre for Radio Astronomy Research, Curtin 
	University, GPO Box U1987, Perth, WA 6845, Australia}

\begin{document}

   \date{Received -- -- --; accepted -- -- --}

\pagerange{\pageref{firstpage}--\pageref{lastpage}} \pubyear{2002}

\maketitle

\label{firstpage}

\begin{abstract}

We observed the supernova remnant (SNR) Puppis A in the 21 cm line with the 
Australia Telescope Compact Array with the aim of determining the systemic 
velocity and, hence, the corresponding kinematic distance. For the compact, 
background sources in the field, we obtain absorption spectra by applying two 
methods: (a) subtracting profiles on- and off-source towards continuum emission,
and (b) filtering short spacial frequencies in the Fourier plane to remove 
large scale emission. One of the brightest features to the East of the shell of 
Puppis A was found to be a background source, probably extragalactic. Removing 
the contribution from this and the previously known unrelated sources, the 
systemic velocity of Puppis A turns out to be limited between 8 and 12 km 
s$^{-1}$, which places this source at a distance of 1.3 $\pm$ 0.3 kpc. From the 
combined images that include both single dish and interferometric data, we 
analyze the distribution of the interstellar hydrogen. We suggest that an 
ellipsoidal ring at $v \sim +8$ km s$^{-1}$ could be the relic of a bubble 
blown by the progenitor of Puppis A, provided the distance is $\lesssim 1.2$ 
kpc. The main consequences of the new systemic velocity and distance as 
compared with previous publications ($v = + 16$ km s$^{-1}$ and $d = 2.2$ kpc) 
are the absence of a dense interacting cloud to the East to explain the 
morphology, and the decrease of the shell size and the neutron star velocity, 
which are now in better agreement with statistical values.

\end{abstract}


\section{Introduction}\label{Int}

H{\sc i} absorption studies towards the continuum background offered by 
supernova remnants (SNR) reveal the distribution of the insterstellar medium 
(ISM) gas along the line of sight and, combined with Galactic rotation models,
can set limits on the remnants' distance. Reliable distance estimates are 
important to determine intrinsic properties of a SNR such as size, age, 
explosion energy or expansion velocity. In addition, H{\sc i} emission at 
velocities beyond a SNR's systemic velocity or away from the direction of the
remnant also give information about the ISM distribution and Galactic structure.
Therefore, H{\sc i} observations help to construct a three dimensional picture 
of the ambient medium where a SNR is evolving, as well as to identify 
foreground or background structures in the Galaxy.

The southern Galactic SNR Puppis A is an extended, distorted shell, $\sim 
50^\prime$ in diameter, estimated to be between 3,700 \citep{Wink+1988} and 
5,200 \citep{Becker+2012} yrs old. It is one of the brightest SNRs in X-rays
\citep[e.g.][]{hwang+08} and has recently been detected in $\gamma$-rays
\citep{Fermi+12}. A pulsating X-ray compact central object (CCO) inside Puppis 
A confirms that the progenitor was a high-mass star \citep{PBW96,ZTP99}. 
\citet{wp2007} showed that the CCO is moving away from the explosion centre 
inferred from optical filaments \citep{Wink+1988}. \citet{EMR+03} found an 
elongated minimum in the H{\sc i} emission at +16 km s$^{-1}$ coincident with 
the path followed by the CCO. 

Several suggestions of interaction between Puppis A and nearby clouds have been 
reported, although the picture is not completely clear yet. \citet{gd+ma88} 
observed Puppis A in the 21 cm line and in the CO (J=1--0) 2.6 mm line and 
found a molecular cloud to the East of the SNR coincident with a flattening in 
the radio continuum shell. The authors interpret this coincidence as evidence 
of a molecular cloud-SNR interaction, and set the systemic velocity of the 
remnant to be around +16 km s$^{-1}$. A subsequent H{\sc i} study using VLA 
observations \citep{EMR+95} supports the same result based on morphological 
coincidences, and infers a distance of 2.2 kpc. However, indubitable tracers of 
interactions between SNR and external clouds, like molecular broadenings or OH 
masers at 1720 MHz, have never been observed. A high resolution CO study 
\citep{paron+08} failed to detect any gas concentration associated with an 
X-ray bright knot to the East of the shell, previously interpreted as coming 
from a shocked interstellar cloud \citep{BEK+05}.

\citet{frail+96} reported single dish observations at 1720 MHz towards
Puppis A performed with the Parkes and Greenbank telescopes, with negative
detections. \citet{beate+00} observed several pointings towards Puppis A
and in the immediate vicinity using the 26-m antenna at the Hartebeestoek Radio
Observatory in South Africa in the four 18-cm lines of OH and concluded that
the systemic velocity of the remnant is 7.6 km s$^{-1}$ rather than +14 km
s$^{-1}$, as proposed based on H{\sc i} \citep{EMR+95} and CO \citep{gd+ma88}
observations. The authors drew attention on the fact that they did detect the 
1720 MHz line in emission, albeit at $\sim 3 $ km s$^{-1}$, where the other 
three lines appear in absorption. This emission may be hinting at the presence 
of an anomalous OH cloud like those reported by \citet{turner82}, which are 
associated with giant clouds and are tracers of the spiral arms. 

The neutral and molecular density distribution around Puppis A is necessary to
model the $\gamma$-ray emission and explain its origin. Additionally, the 
hydrogen column density is a key parameter to interpret the X-ray emission. In 
all cases, it is essential to establish the SNR's systemic velocity. Although 
morphological arguments help to infer this velocity, the most reliable method is
to analyze the H{\sc i} absorption. We note that high quality absorption studies
towards Puppis A are lacking. The previous VLA H{\sc i} study \citep{EMR+95} 
has a poor velocity resolution (5.2 km s$^{-1}$) and several residual sidelobes 
make the identification of emission or absorption features rather unclear. The 
other absorption study conducted by \citet{beate+00} in OH lines has poor 
angular resolution ($\sim 20^\prime$) and sampling (only 11 pointings on and 
off the SNR). In this paper we analyze new high angular and spectral resolution 
data in the H{\sc i} 21-cm line in a mosaic centered at Puppis A in order to 
shed light on its kinematic distance and on the gas density distribution into 
which the shock front is expanding. 

\section{Observations and data reduction}\label{Obs}

\begin{figure*}
\centering
\includegraphics[width=18cm]{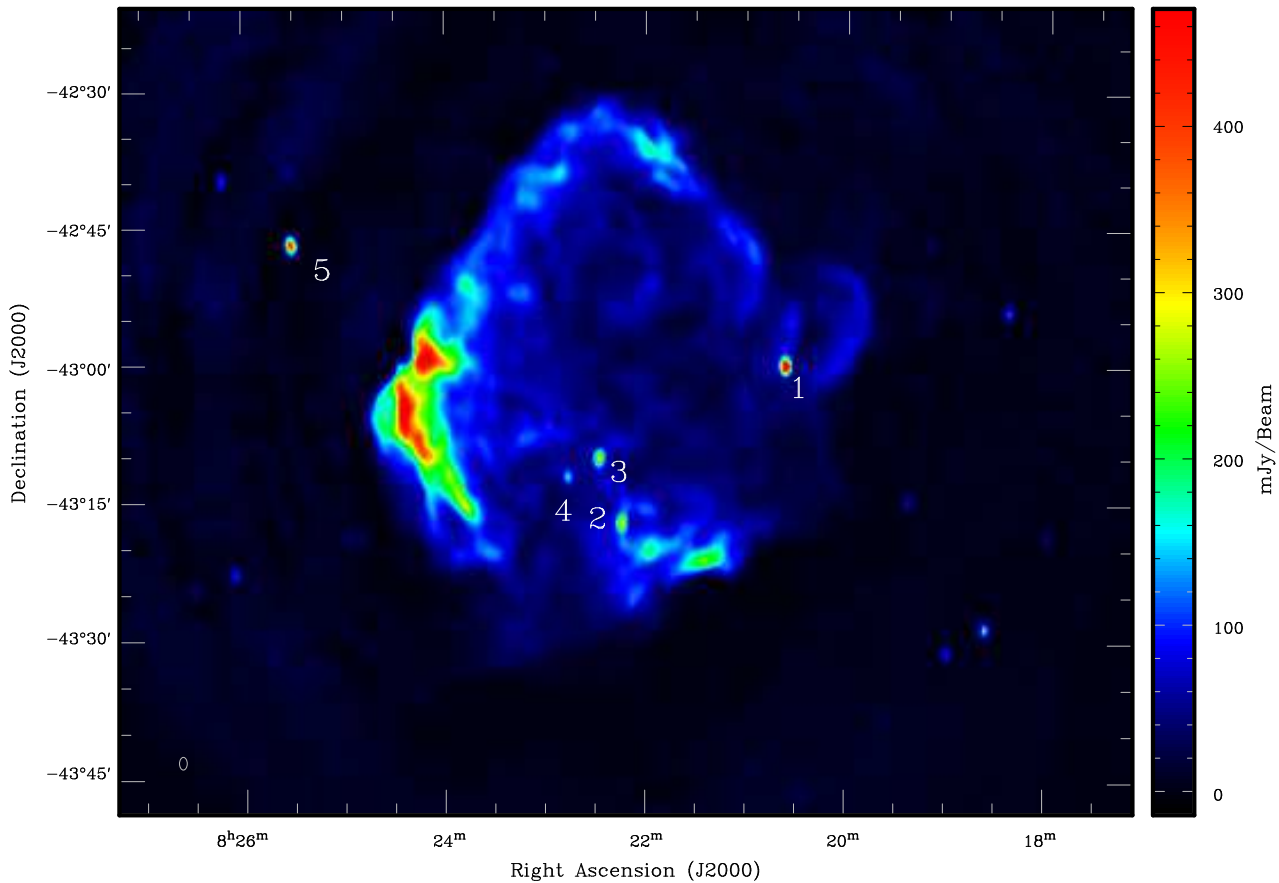}	
\caption{A continuum image of Puppis A at 1.4 GHz reproduced from \citet{pap1}. 
The beam size is 82.2 $\times$ 50.6 arcsec, with a position angle of 
-0\hbox{$.\!\!{}^\circ$}55 degrees, and is plotted at the bottom left corner. 
The flux density scale is shown at the right. The noise level is 1.5 mJy 
beam$^{-1}$. The white numbers 1 to 5 on selected compact sources correspond to 
the labelling applied in \citet{pap1}. A color version is available in the 
on-line paper.} 
     \label{ContMos}
\end{figure*}

Puppis A was observed simultaneously in the H{\sc i} 21 cm line, in the four OH 
18 cm lines, and in radio-continuum with the Compact Array Broadband Backend 
(CABB) of the ATCA in two 13 hr runs.
The first run took place on 2012 May 20 using the array configuration EW 352 
(baselines varying from 31 to 352 m excluding the 6th antenna), and the second, 
on 2012 July 6 in the 750A configuration (baselines from 77 to 735 m). A 
thorough analysis of the continuum data was published in \citet{pap1}. The 
primary flux and bandpass calibrator was PKS 1934-638, while PKS 0823-500 was 
used for phase calibration. Due to the large angular size of Puppis A, 
observations were made in mosaicking mode with 24 pointings in order to cover 
not only the SNR but also its surroundings. Following the Nyquist theorem, the 
pointings were separated by 19.6 arcmin to optimize the sampling at 1.4 GHz. 
The radio continuum was observed using a 2 GHz bandwidth divided into 2048 
channels of 1 MHz width, centred at 1750 MHz. The CABB allowed us to observe 
simultaneously a fine resolution 1-MHz 'zoom' band composed of 2048 channels of 
0.5-kHz width, which were centered on the H{\sc i} line and on the four ground 
state OH lines, at 1612, 1665, 1667 and 1720 MHz.

The data reduction was carried out with the {\sc miriad} software package 
\citep{Sault+95}. The first run observations (configuration EW352) were
affected by a bug that was fixed before the second run took place: the CABB
only read the first 8 characters of a mosaic field name, and as a result,
fields 10 to 19 were set to the phase center of field 1, and fields 20 to 24,
to that of field 2. To overcome this problem, we have split the uv data for the
different fields, set the correct phase center for each field in its header,
and shifted the data to the new phase center using the {\sc miriad} task
UVEDIT. Having performed these steps, we followed henceforth a joint
deconvolution. This correction was not necessary for the second run
observations (configuration 750A).

Apart from the long mosaic name length bug, there was an additional CABB bug
that affected the observations made in the 1 MHz zoom mode: the first correlator
cycle was corrupted in such a way that the emission was shifted in frequency.
To fix this problem, all first cycles had to be discarded. However, since
off-source cycles are discarded by default when loading the data into {\sc
miriad}, in principle it is not possible to determine when the first cycle in a
scan is the first correlator cycle.  Thus, we loaded the data with the 'unflag'
option so as to keep all data, even the flagged ones, and then flagged every
first cycle using the {\sc miriad} task QVACK. This procedure was applied for
data from both May and July observe runs.

The remaining data reduction was standard. To improve the signal-to-noise
ratio and save disk space at the same time, the data were Hanning smoothed as
they were loaded into {\sc miriad}, and only every second channel was kept,
amounting to a total of 1536 channels of 1 kHz width. After amplitude, bandpass 
and gain calibration, the continuum component was removed from the uv-data 
subtracting a linear fit to 550 line-free channels. An H{\sc i} cube was 
constructed merging the data obtained with both ATCA configurations. Uniform 
weighting was applied.

The noise level was reduced at the expense of reducing the spectral resolution:
we collapsed each four channels into one so as to increase the number of
visibilities per channel by a factor of four, and kept only one in every four 
channels. As a result, the H{\sc i} cube has 213 channels spanning from -23 km 
s$^{-1}$ to 151.81 km s$^{-1}$, with an increment of 0.82 km s$^{-1}$. Although 
MOSMEM is a more robust task to clean a mosaic, it only copes with positive 
flux density values. Since H{\sc i} absorption data are negative, we performed 
the cleaning with the {\sc miriad} task MOSSDI which, although optimized for 
point sources, can deal with negative components. After a deep cleaning, most 
sidelobes were removed. The noise achieved varies between approximately 2 and 9 
mJy beam$^{-1}$. The noise is lower for channels with lower H{\sc i} emission. 
Besides, regions that are covered by more than one mosaic field have a lower 
noise, which means that the noise increases towards the edges of the map. The 
images were restored with a beam of 118.3 $\times$ 88.9 arcsec, with a position 
angle of -4\hbox{$.\!\!{}^\circ$}3.

To recover structures with the shortest spatial frequencies, the ATCA data were
combined in the uv plane with single-dish data from the 64-m Parkes telescope,
which are part of the Galactic All-Sky Survey (GASS). This step was performed
with the {\sc miriad} task IMMERGE. No tapering was applied to the
low-resolution cube. All pixels falling outside the mosaic were blanked out.

As an alternative way to produce absorption profiles, we have also followed the 
method proposed by \citet{dickey+03}: we constructed a higher resolution cube 
by filtering out all spatial frequencies lower than 168m (0.8 k$\lambda$). 
We have not subtracted any continuum baseline this time, and the line free 
channels were used later to form an ancillary radio continuum image. After 
cleaning the H{\sc i} cube with the {\sc miriad} task MOSSDI, a noise level of 
$\ \hbox{$_\sim\!\!\!\!{}^>$}\ 30$ mJy beam$^{-1}$ was achieved. The beam is 
69.4 $\times$ 45.6 arcsec, with a position angle of -1\hbox{$.\!\!{}^\circ$}9. 
The way in which this cube will be used to obtain absorption spectra will be 
explained in detail in Section \ref{abs}.

\section{Results}

\subsection{Absorption profiles}\label{abs}

Puppis A is an extended radio source, with a diameter of almost 1-degree.
Therefore, the density distribution of the interstellar medium (ISM) is not 
expected to be uniform across its face. In order to avoid misinterpretations 
of absorption profiles towards Puppis A that could arise in local 
inhomogenities, we will make use of the strong compact sources 1 -- 5 (Fig.
\ref{ContMos}) to compare their profiles with Puppis A and attempt set a 
reliable limit in the systemic velocity of the SNR.  

The equation of radiative transfer states that the emission $T_{{\rm b}_v}$ at 
a velocity $v$ measured by a radiotelescope is\\
\begin{equation}\hskip 2 cm
T_{{\rm b}_v} = T_s (1-e^{- \tau _v}) + T_c \, e^{-\tau _v} ,
\label{transp1}
\end{equation}
where $T_s$ and $\tau _v$ are the spin temperature and the optical 
\begin{landscape}
\begin{figure}
\hfill
\begin{minipage}{0.25\textwidth}
\includegraphics[width=\textwidth]{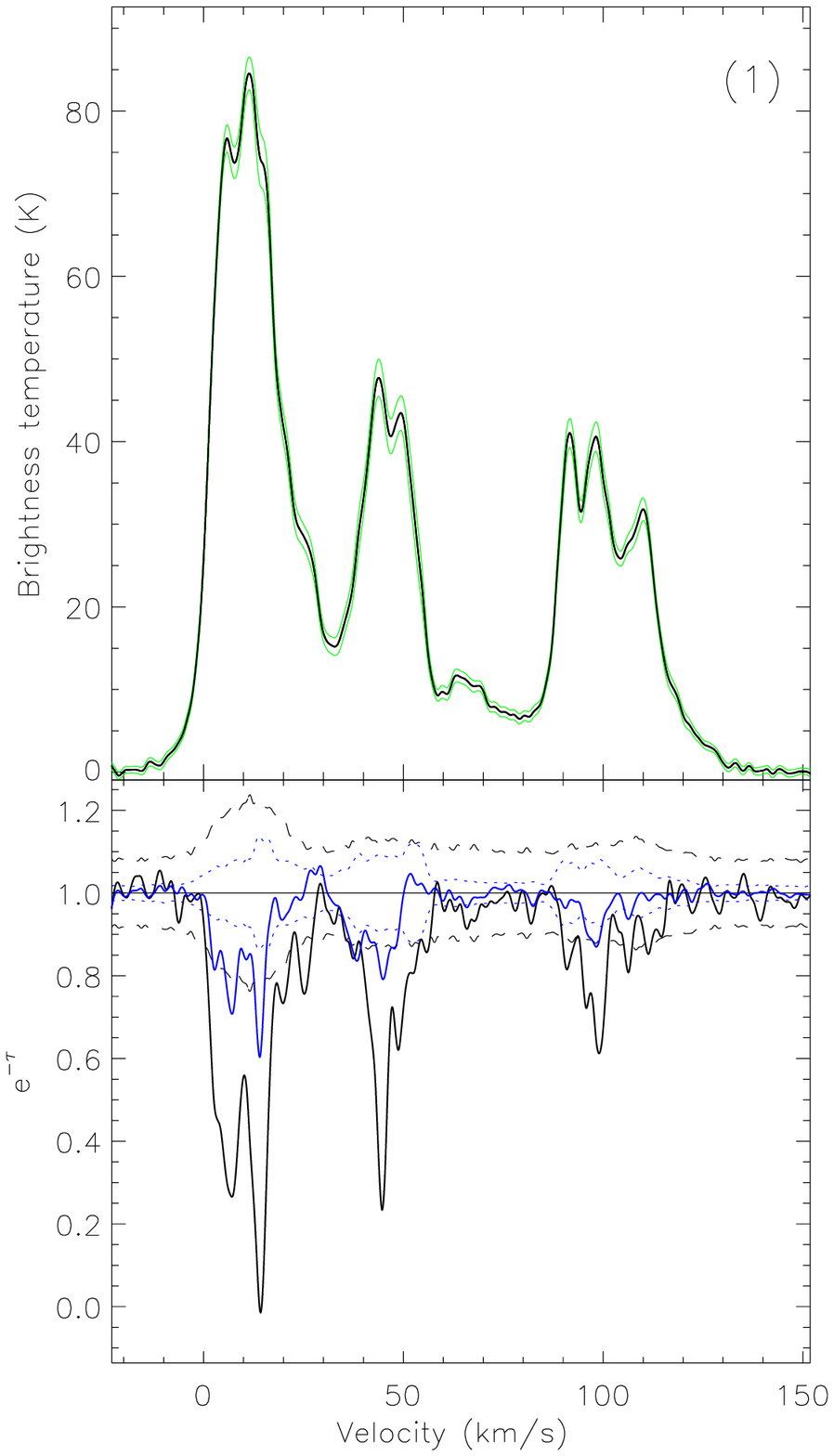}
\end{minipage}
\hskip 0.25cm
    \hfill
    \begin{minipage}{0.25\textwidth}
\includegraphics[width=\textwidth]{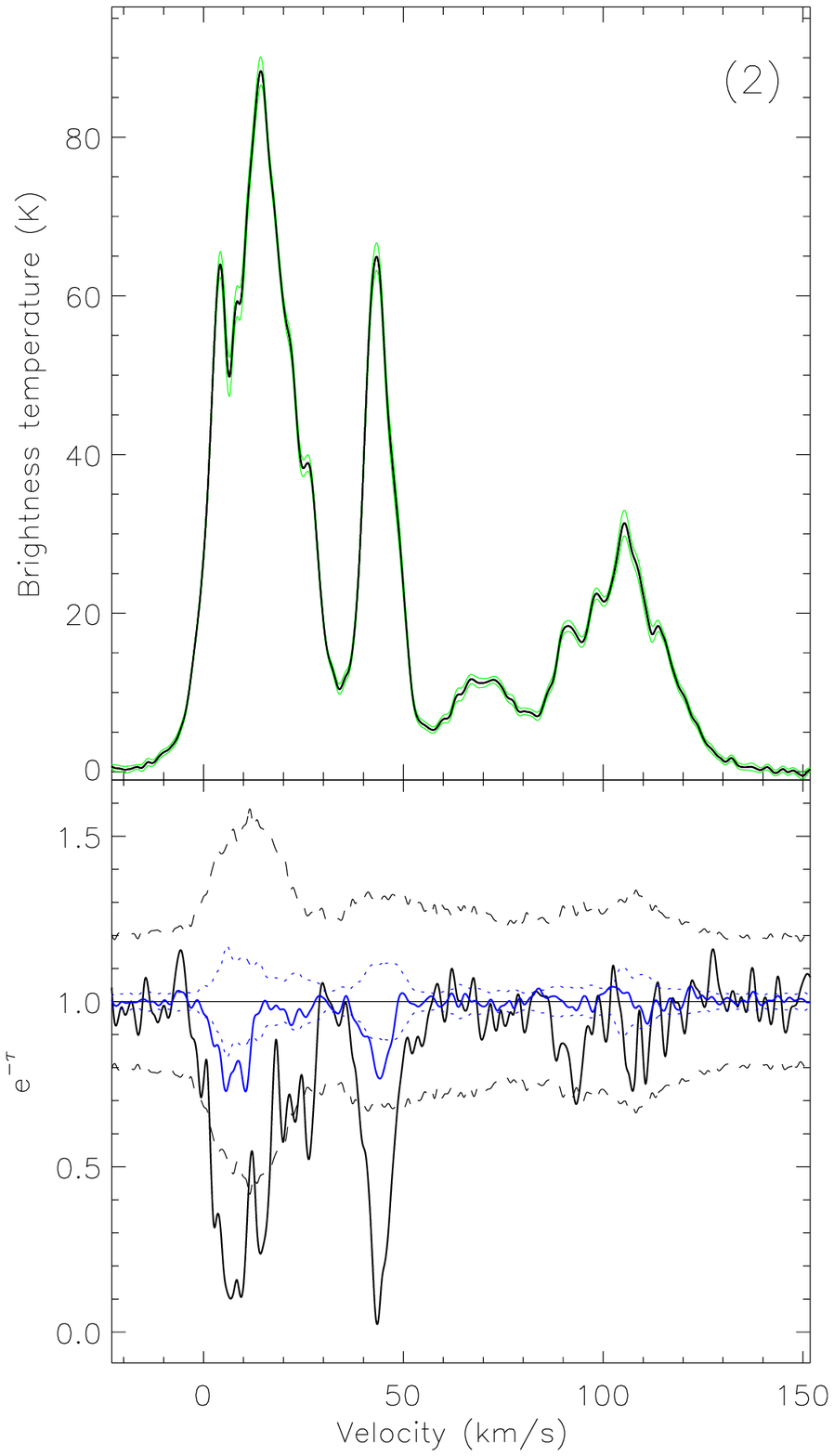}
\end{minipage}
\hskip 0.25cm
\hfill
    \begin{minipage}{0.25\textwidth}
\includegraphics[width=\textwidth]{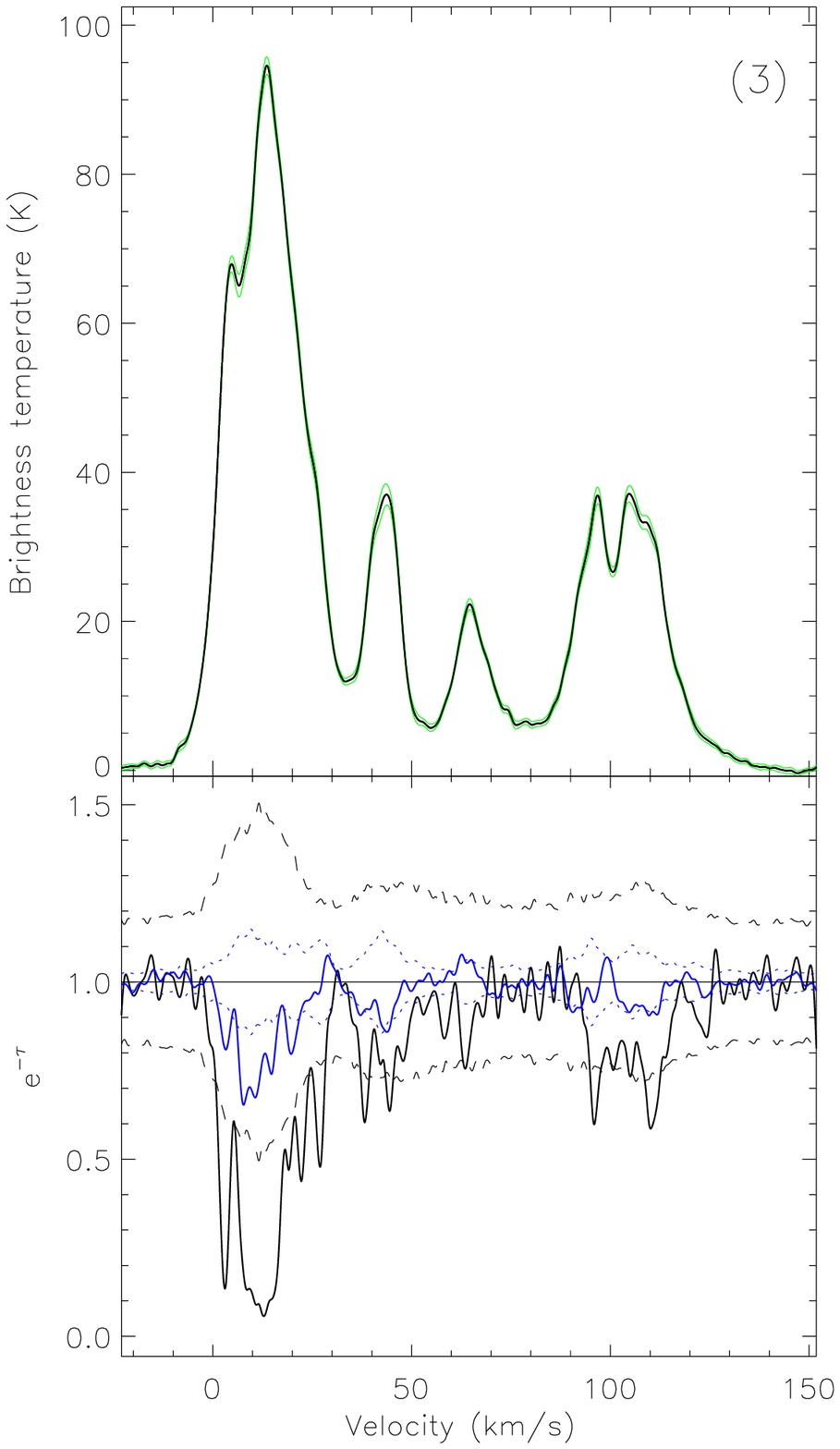}
\end{minipage}
\hskip 0.25cm
    \hfill
    \begin{minipage}{0.25\textwidth}
\includegraphics[width=\textwidth]{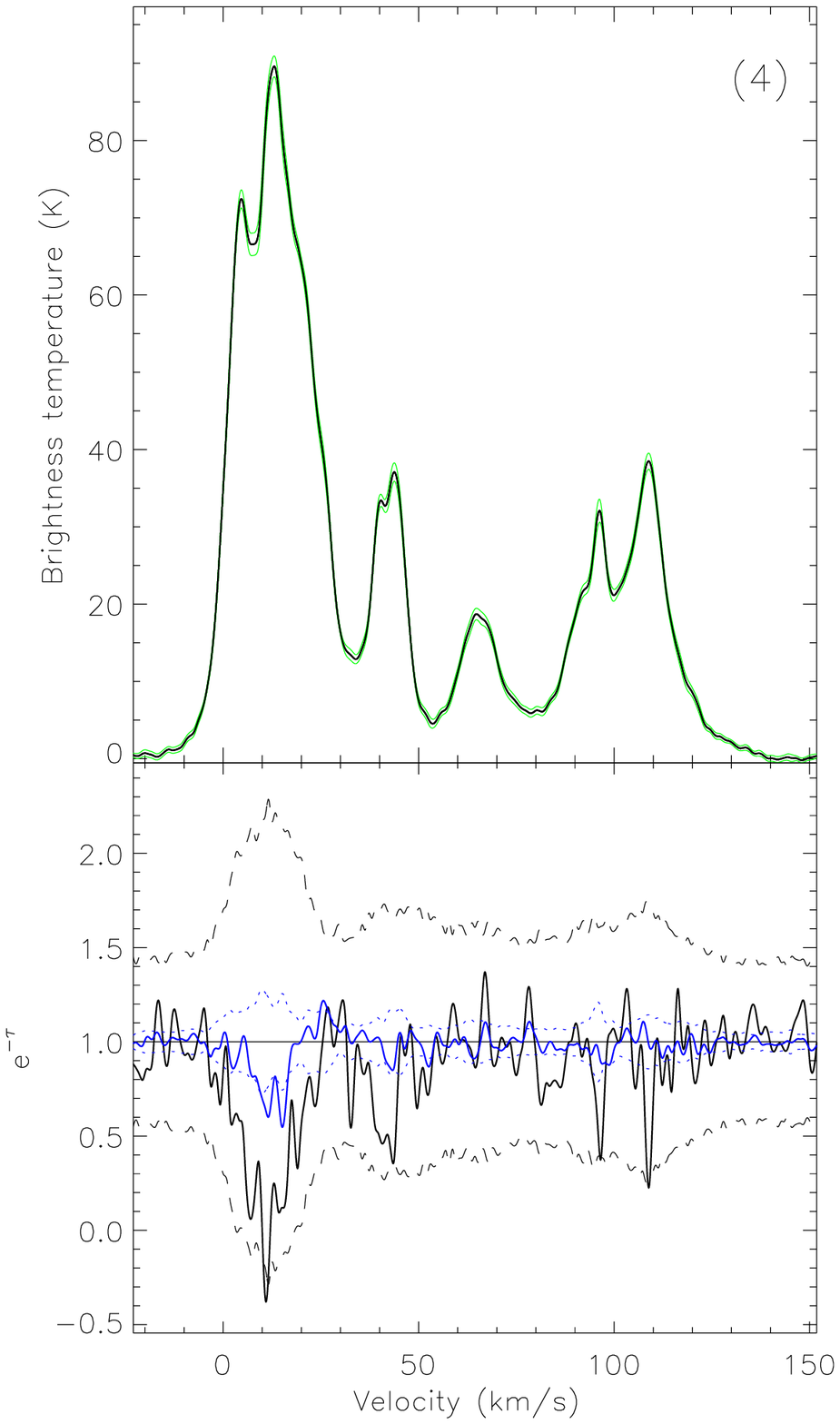}
\end{minipage}
\hskip 0.25cm
\hfill
    \begin{minipage}{0.25\textwidth}
\includegraphics[width=\textwidth]{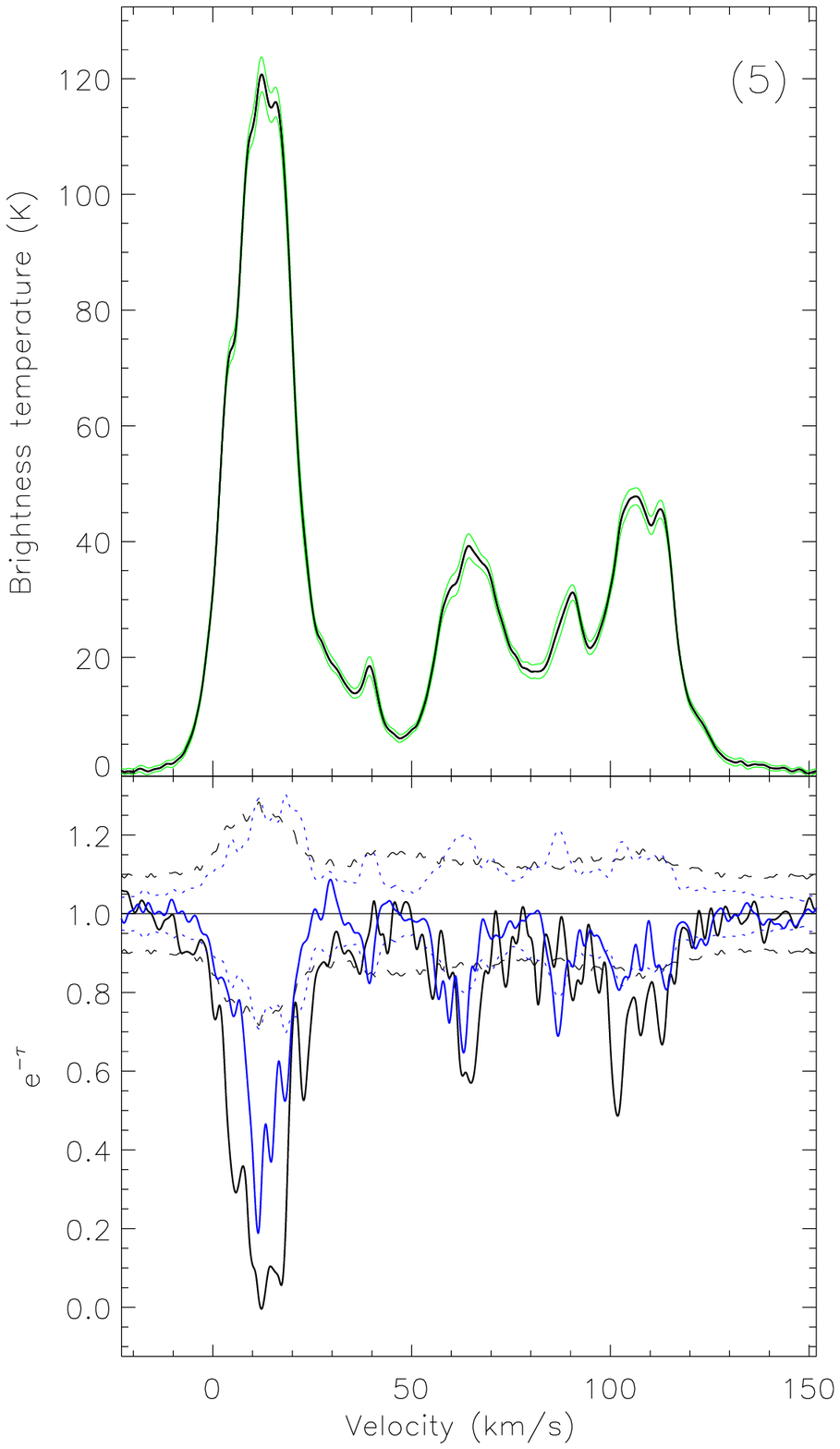}
\end{minipage}
\caption{Emission-absorption spectrum pairs for the five compact sources shown 
in Fig. \ref{ContMos}. The source number is shown at the top right corner of
each frame. The expected profiles are shown at the top, with the error
envelopes in green. At the bottom, absorption profiles computed with two 
different methods are displayed: through the subtraction $T_{on} - T_{off}$
(blue solid lines), and through a short wavelength uv-filtering weighted by 
the radio continuum (black solid lines), as explained in the text. The error 
envelopes corresponding to each method are shown as dotted (first method) and
dashed (second method) lines following the same color code. In all cases, a 
Hanning smoothing was applied.}
\label{perfs-cs}
\end{figure}

\begin{table}
\caption{Parameters of compact sources}
\label{cat-cs}
\centering
\begin{tabular}{cccccccc}
\hline\hline
Source& RA (J2000) & Dec. (J2000) & ~~~Spectral & ~~~~Spectral &\ \ Cutoff levels for 
unfiltered \ \ &\ \ Cutoff level for uv-filtered\ \ & Peak radio continuum\\
number& &&~~~index$^a$ &~~~index$^b$ &radio continuum template $^c$ & radio continuum 
template $^d$& brightness temperature \\
& ~h~ m ~ s ~& ~~$^\circ$ ~~~$^\prime$ ~~~$^{\prime\prime}$ & & & ~~ $T_{cut_{1}}$~(K) ~~~~ $T_{cut_{2}}$~(K) ~~~~ & $T_{cut}$~(K) ~~~~ & ~~~~~~(K) ~~~~~\\
\hline 
1 &08 20 35.6 & -43 00 23& -0.76$\pm$0.04 &-0.86$\pm$0.04&10.8 ~~~~~~~~~~~~ 23.2 ~~~~~& 29.2 ~~~~~~& 106.3\\
2 &08 22 13.4 & -43 17 38& -0.91$\pm$0.07 &-1.09$\pm$0.05&17.8 ~~~~~~~~~~~~ 20.9 ~~~~~& 17.4 ~~~~~~& 43.5 \\
3 &08 22 27.0 & -43 10 29& -0.26$\pm$0.07 &-0.56$\pm$0.5 &10.8 ~~~~~~~~~~~~ 12.4 ~~~~~& 16.6 ~~~~~~& 47.3 \\
4 &08 22 45.7 & -43 12 38& -0.75$\pm$0.15 &-0.94$\pm$0.08&~8.8 ~~~~~~~~~~~~ 12.2 ~~~~~& ~8.5 ~~~~~~& 24.6\\
5 &08 25 30.7 & -42 46 57& -0.92$\pm$0.04 &-1.09$\pm$0.09&~0.8 ~~~~~~~~~~~~~ 3.1 ~~~~~& 21.0 ~~~~~~& 84.2\\
\hline
\end{tabular}
\begin{list}{}{}
\item{$^{\rm a}$} Index based on T-T plot \citep{pap1}. 
\item{$^{\rm b}$} Index based on slope of $S(\nu)$ versus $\nu$ \citep{pap1}.
\item{$^{\rm c}$} Pixels above $T_{cut_{1}}$ were replaced by bilinear 
interpolation, and pixels above $T_{cut_{2}}$ were averaged to compute the 
absorption profile. 
\item{$^{\rm d}$} Pixels above $T_{cut}$ were averaged to compute the 
absorption profile. 
\end{list}

\end{table}

\end{landscape}
\noindent 
depth of the column of H{\sc i} gas, and $T_c$ is the background continuum 
emission. The continuum emission was subtracted from the line data as described 
in Sect. \ref{Obs}, hence Eq. \ref{transp1} becomes
\begin{equation}\hskip 2 cm
T_{{\rm L}_v} = T_{{\rm b}_v} - T_c = (T_s - T_c) (1-e^{- \tau _v}).
\label{transp2}
\end{equation}

To construct absorption profiles, we need to infer what the profile would be if 
the continuum source were not present. This profile, usually called $T_{off}$, 
must be subtracted from the profile measured on the continuum source ($T_{on}$).
According to Eq. \ref{transp2}, the difference $T_{on} - T_{off}$ yields
\begin{equation}\hskip 2 cm
T_{on} - T_{off} = - T_c \, (1 - e^{- \tau _v} )
\label{transp3}
\end{equation}
\noindent or equivalently,
\begin{equation}\hskip 2 cm
e^{- \tau _v} = { {T_{on} - T_{off} + T_c} \over T_c} .
\label{transp4}
\end{equation} 
\noindent The method we applied to estimate $T_{off}$ is as follows: we 
extracted a box surrounding each compact source allowing for a large enough 
area around it as to include a statistically significant number of pixels free 
of continuum emission. The continuum image was used as a template, and all 
those pixels with flux densities above a cutoff level low enough as to exclude 
any continuum  emission from the compact source ($T_{{cut}_1}$) were blanked 
in each H{\sc i} channel. The remaining pixels were used to fit a bilinear 
function to the box in each channel, and this fit was employed to fill in the 
blanked areas.  With this procedure, the pixels of the H{\sc i} cube above a 
cutoff continuum level in the template ($T_{{cut}_2}$) yield the on-source 
profile when the original values are used and the expected off-source profile 
when these are replaced by the bilinear fit. The error of the interpolation is 
given by the rms of the residuals between the bilinear approximation and the 
original values for the pixels used to perform the fit.

\begin{figure}
\centering
     \label{pup_east}
\begin{tabular}{c}
\includegraphics[width=0.45\textwidth]{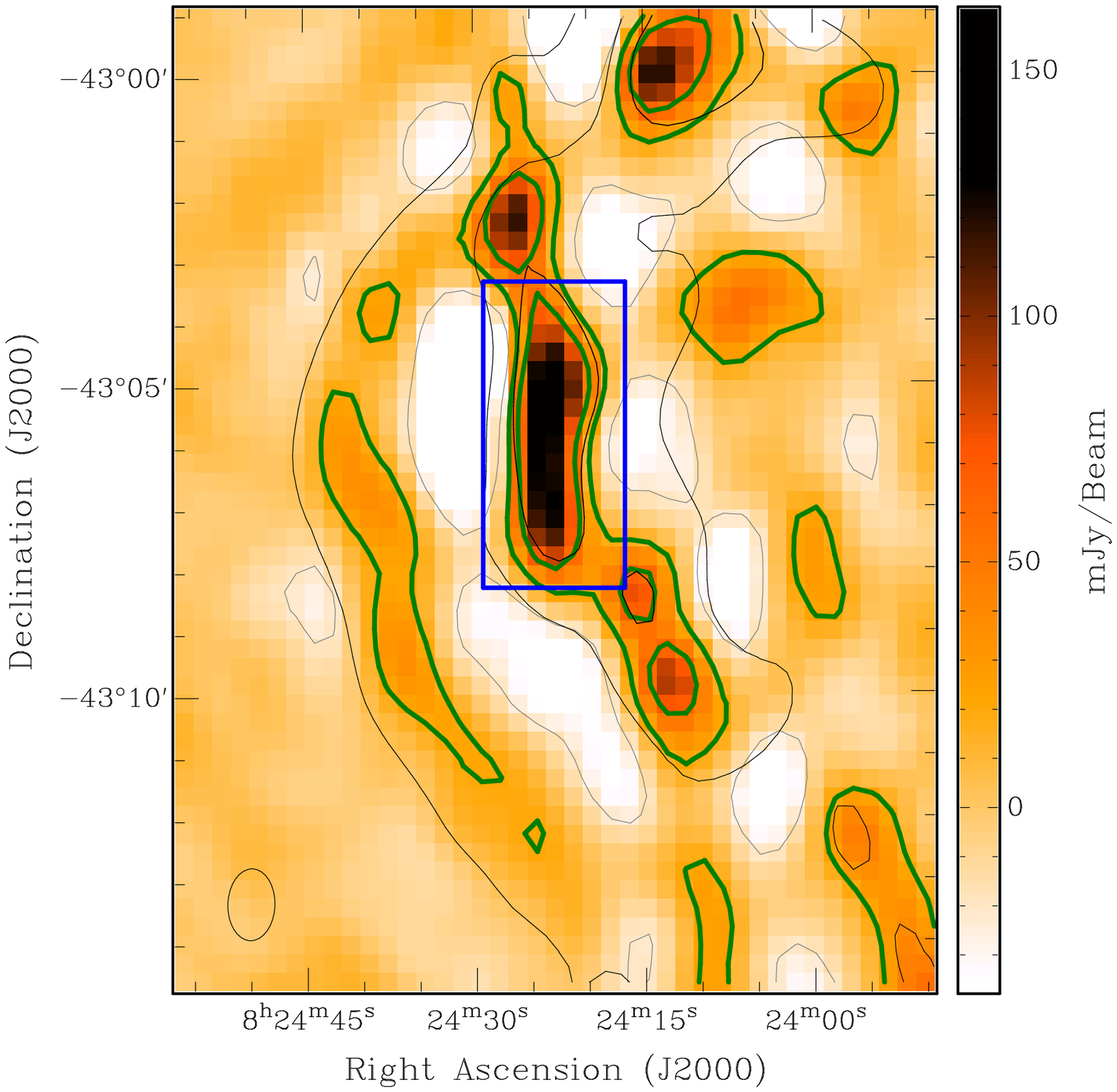}\\{\ } \\
\includegraphics[width=0.39\textwidth]{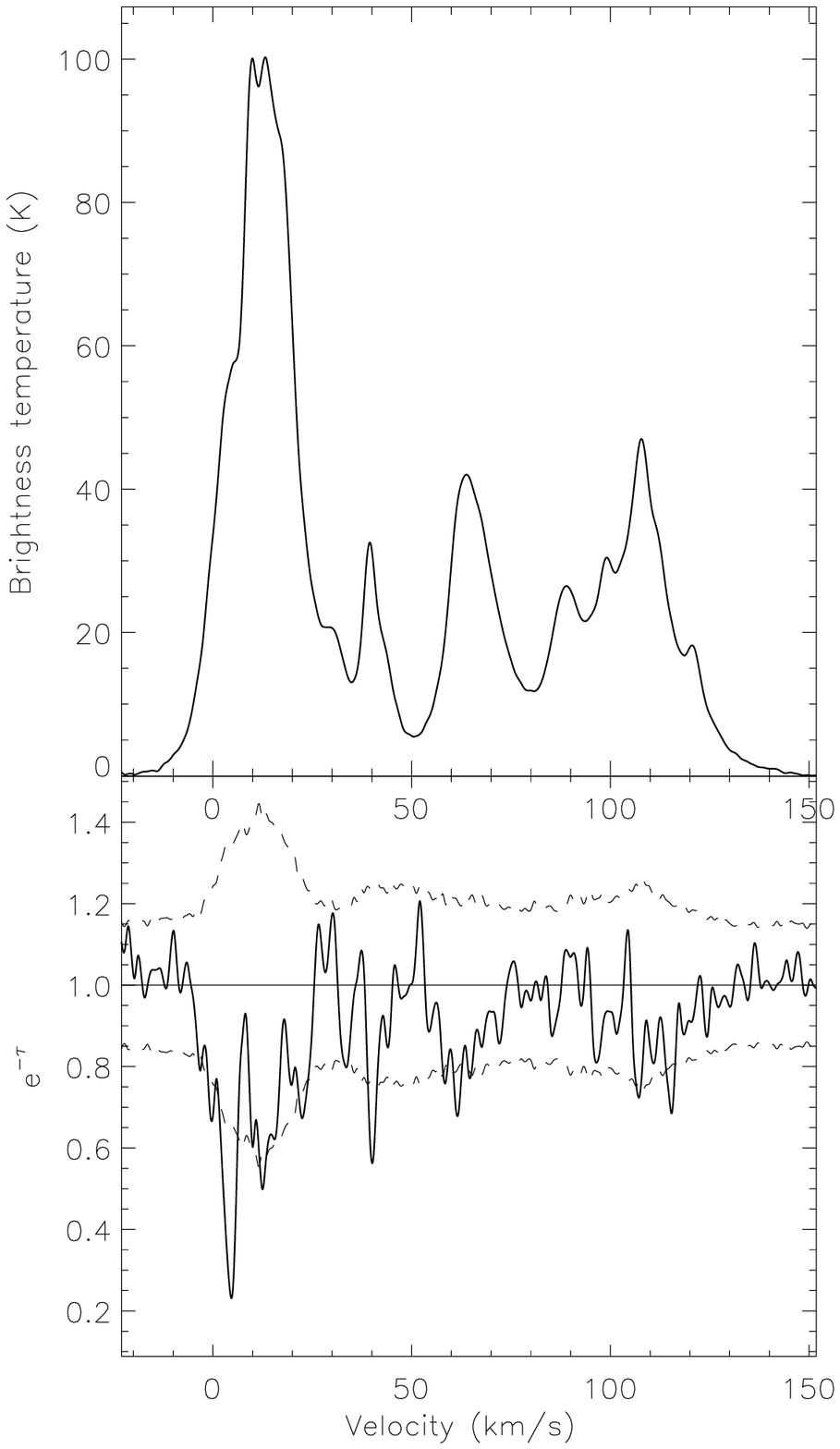}
\end{tabular}
\caption{{\it Top:} Radio continuum image of the eastern region of Puppis A 
as obtained from the line-free channels of the uv-filtered cube at 1420 MHz. 
Thick green contours at 20 and 60 mJy beam$^{-1}$ and thin grey contours at 
-20 mJy beam$^{-1}$ are overlaid. The beam is shown at the bottom left corner. 
To facilitate the description, a few black contours from the unfiltered 
continuum image at 1.4 GHz are also plotted. {\it Bottom:} Emission-absorption 
spectrum pair towards the region of Puppis A enclosed by the green contour at 
60 mJy beam$^{-1}$ within the inner blue box (top panel). The 3$\sigma$-error 
envelope is plotted with a dashed line. A Hanning smoothing was applied.}
\end{figure}

The emission-absorption spectrum pairs for the five compact sources indicated 
in Fig. \ref{ContMos} are plotted in Fig. \ref{perfs-cs}. For each, the upper 
frame is the expected profile (black solid line) with the error envelope,
computed as described above, in green. The lower frame displays the absorption 
profile (blue solid lines) estimated using Eq. \ref{transp4}, with the 
corresponding error envelope in blue dashed lines.

For extended areas, the bilinear aproximation is less accurate since the gaps
between valid data to compute the fit are too large and a linear interpolation
may not represent the missing points. An alternative way to estimate absorption
profiles is to filter out extended emission in the Fourier domain. As mentioned 
in Sect. \ref{Obs}, we have constructed an H{\sc i} cube without subtracting 
the continuum level and using only visibilities above 0.8 k$\lambda$, so that 
features with sizes larger than $\sim 4.5$ arcmin are filtered out. Hence, the 
term involving $T_s$ in Eq. \ref{transp1} can be neglected and the remaining 
terms can be re-written as $e^{- \tau _v} = {T_{{\rm b}_v} / T_c}$. To enhance 
absorption features in the profiles, we further weighted each pixel, 
$T_{{\rm b}_v}$, in the filtered H{\sc i} cube by the corresponding radio 
continuum flux. 
To estimate the errors, for each channel we computed the rms on selected boxes 
towards three different regions away from radio continuum sources and obtained 
their averages. The profiles obtained by this method are plotted in black solid 
lines, with the 3$\sigma$-error envelopes as black dashed lines.  

\begin{figure}
\centering
\includegraphics[width=0.5\textwidth]{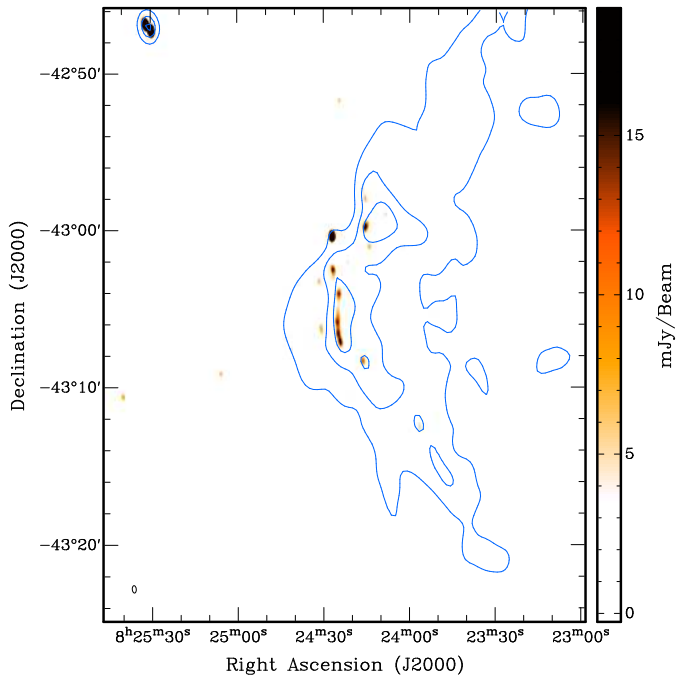}	
\caption{Image of the eastern region of Puppis A at 2.4 GHz filtered in the 
Fourier domain between 3 and 7 k$\lambda$. The beam  is 28.1 $\times$ 15.2 
arcsec, with a position angle of 4\hbox{$.\!\!{}^\circ$}7 degrees, and is 
plotted at the bottom left corner. The flux density scale is shown at the 
right. The noise level is 2 mJy beam$^{-1}$. To facilitate the comparison
with the shell, a few blue contours from the unfiltered continuum image at 
1.4 GHz are overlaid.}
     \label{extragal}
\end{figure}

In Table \ref{cat-cs} we list the relevant parameters for the five compact 
sources and the determination of their absorption profiles. The first three 
columns contain their labels and coordinates. In the next two columns, we 
reproduce the spectral indices as measured in \citet{pap1} following two 
methods: T-T plots (4$^{\rm -th}$ column) and a linear fit to the logarithmic 
plot $S_\nu$ versus $\nu$ (5$^{\rm -th}$ column). The 6$^{\rm -th}$ column 
lists the cutoff brightness temperature of the radio continuum template map 
above which pixels in the H{\sc i} cube were blanked and replaced by the 
bilinear fit computed with the remaining valid pixels within a box around each 
source. The 7$^{\rm -th}$ column displays the cutoff brightness temperatue of 
the radio continuum template map above which the corresponding pixels in the 
H{\sc i} cube were averaged to construct the $T_{on}$ and $T_{off}$ profiles. 
The 8$^{\rm -th}$ column indicates the cutoff brightness temperature of the 
uv-filtered radio continuum template map; the corresponding pixels above this 
value in the uv-filtered H{\sc i} cube were used to construct a weighted 
absorption profile. The last column lists the peak radio continuum brightness 
temperature in the uv-filtered image, and is added with the intention of 
understanding the difference in the quality of the absorption profile of each 
source in terms of the difference in their intensities.

\begin{figure}
\centering
\includegraphics[width=0.5\textwidth]{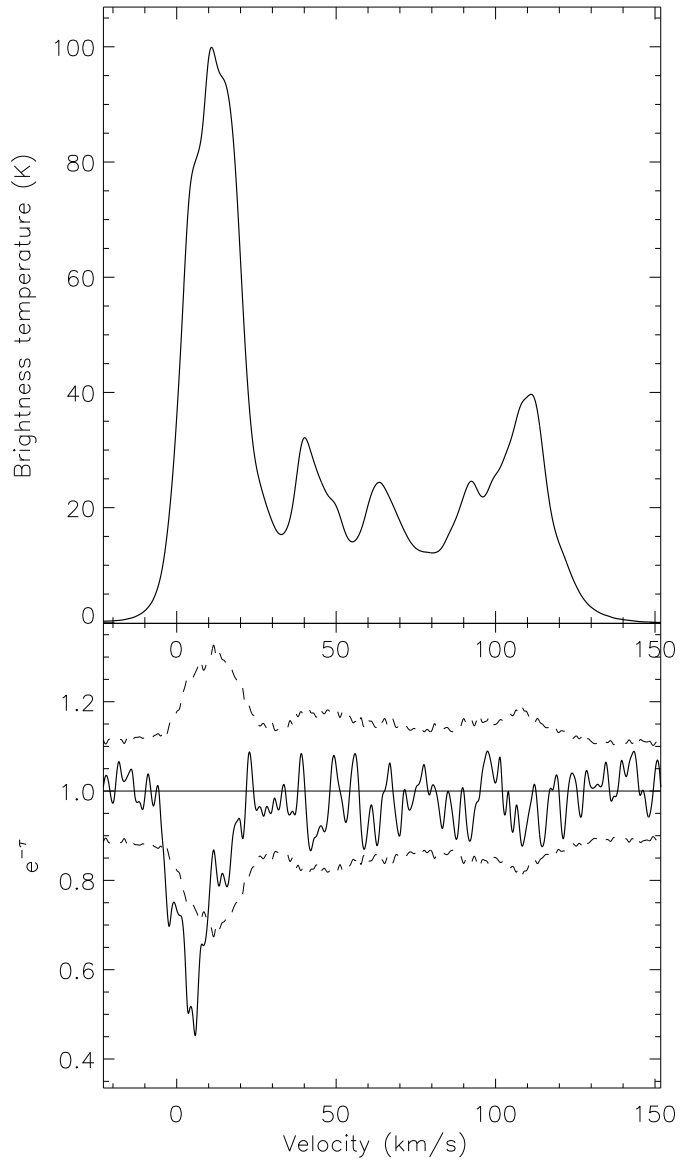}
\caption{Emission-absorption spectrum pair of Puppis A. The compact sources 1
to 4 and the eastern bar have been removed. The 3$\sigma$-error envelope is 
plotted with a dashed line. Hanning smoothing was applied.}
     \label{pup_all}
\end{figure}

\begin{figure*}
\includegraphics[width=16.25cm]{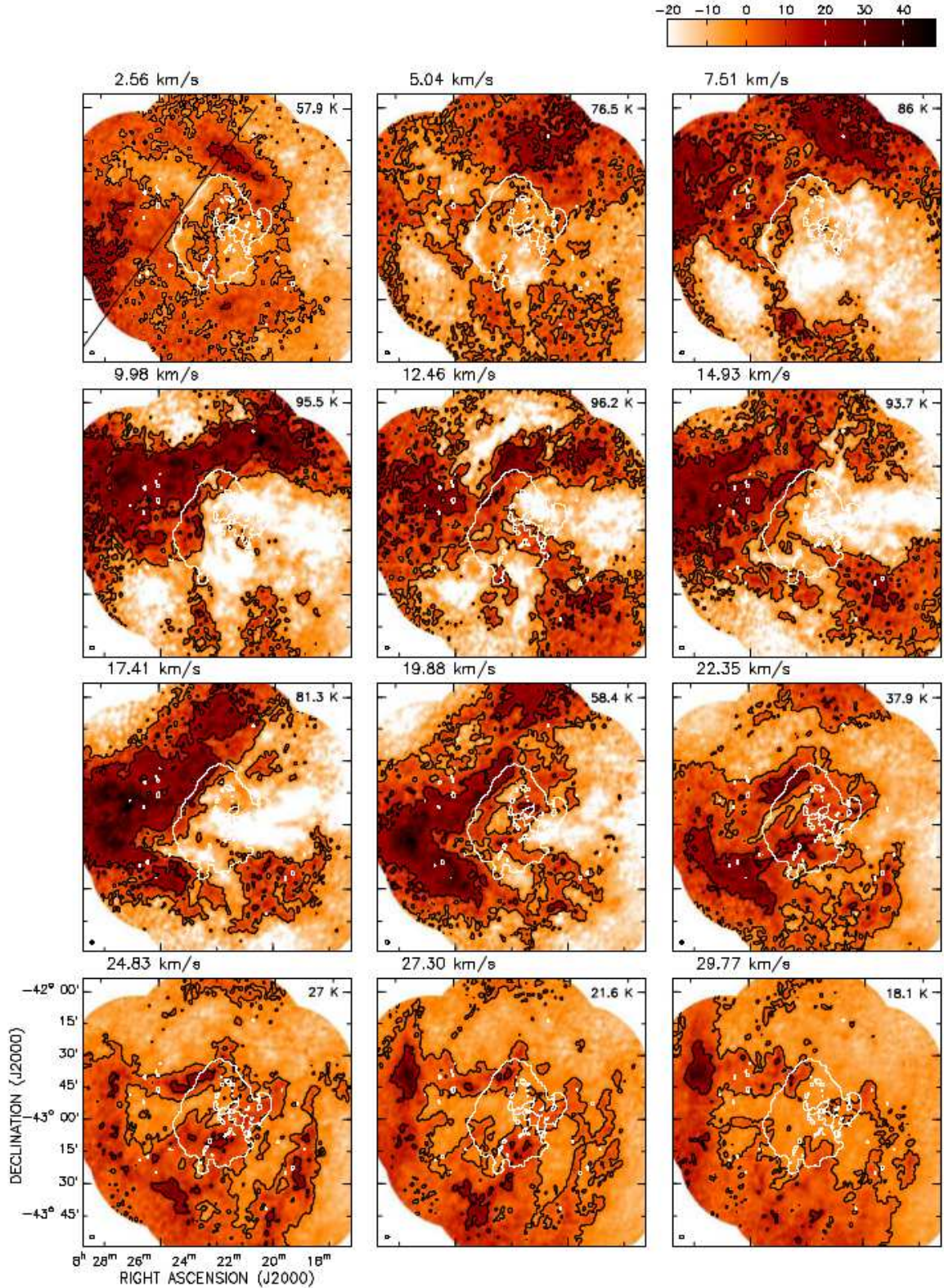}
\caption{H{\sc i} emission distribution between +2.56 and +29.77 km s$^{-1}$. 
The central velocity of each panel is given in the upper left-hand corner. 
Black contours are at 0 and 15 K. The white contour corresponds to the 0.03 mJy 
beam$^{-1}$ level of the radio continuum emission. The tilted line plotted 
superimposed on the emission at +2.56 km s$^{-1}$ indicates the line of 
Galactic latitude $b$ = -3$^{\circ}$. To use a same color scale, an average 
brightness temperature, indicated at the top right corner inside each frame, 
was subtracted from every channel. The color scale, in K, is shown on top of 
the figure.}  
\label{sethi}
\end{figure*}

It is clear that the uv-filtering method produces spectra with deeper, more 
discernible absorption features. Besides, we note that the two methods give
a closer result for source 5, which is the only one that is not projected 
behind Puppis A. The reason could be that since the rest of the sources are 
confused with the diffuse emission from the remnant, it is not easy to obtain 
an H{\sc i} profile $T_{on}$. Therefore, we will apply the uv-filtering method 
to measure the absorption on Puppis A, since it proves to be the most 
convenient tool for that purpose.

To isolate the contribution from Puppis A only in the uv-filtered cube, we 
blanked four boxes around sources 1 to 4. A first spectrum showed several 
absorption features up to v $\gtrsim +100$ km s$^{-1}$, implying that additional
extragalactic sources were contaminating the emission from Puppis A. Thus,
we constructed profiles over selected regions all over the remnant. An
intrinsic problem of the uv-filtering method is that only the smallest features
survive, and the fewer visibilities used translates in a poorer imaging and a
higher noise. Therefore, a limited number of relevant regions are used for the 
absorption analysis. In the map presented in the top frame of Fig. 
\ref{pup_east}, we show the region where the most significant results are
obtained. We constructed an absorption spectrum towards the area enclosed by
the green contour at 60 mJy beam$^{-1}$ within the inner rectangular box. The 
resulting spectrum, shown at the bottom, shows several absorption features at 
different velocities up to the outer Galactic limits. Most features are beyond 
the error band, indicating that the continuum emission arises in a background 
source, probably of extragalactic origin. This is in agreement with the result 
reported by \citet{pap1}, who found that this feature (box 3 in their Fig. 7) 
has a steep spectrum, $\alpha = -0.8 \pm 0.4$. 

To get more insight into this finding, we built a radio continuum image at 2.4 
GHz, filtering all visibilities below 3 k$\lambda$ to minimize the contribution 
from the extended emission associated with Puppis A. The image is shown in Fig. 
\ref{extragal}. A few contours in blue from the continuum unfiltered image 
presented in Fig. \ref{ContMos} are plotted to facilitate the location of the 
newly discovered features within Puppis A. The compact source 5 appears near 
the top left corner of the map. The bright compact source located at the 
outermost contour of the shell, at RA(2000)=8$^{\mathrm h}$ 24$^{\mathrm 
m}$ 27$^{\mathrm s}$, Dec.(2000)=$-43^\circ$ 00\hbox{$^{\prime}$} 
24\hbox{$^{\prime\prime}$} was reported in \citet{pap1} under number 27, as 
well as the two weak sources 30 and 33, both several minutes away from Puppis 
A, at a declination of $\sim -43^\circ$10\hbox{$^{\prime}$}. It was not 
possible to determine the spectral index for any of these sources. In all radio 
continuum images of Puppis A, the brightest features are the extragalactic 
compact source 1 and the eastern filament at RA(2000) $\sim 8^{\mathrm h}$ 
24$^{\mathrm m}$ 25$^{\mathrm s}$. Our uv-filtered map suggests that the 
eastern filament is actually an elongated extragalactic source, roughly 2 
minutes long, which we will label G260.72-3.16. Other point-like compact 
sources are observed within a \ $\sim 5$ minute radius around it.

Given that the bright eastern filament is a background source, we blanked this 
area and obtained a new spectrum towards the whole remnant. A correction based 
on the line free channels was applied to account for the continuum baseline not 
subtracted in the uv-plane.
The result is depicted in Fig. \ref{pup_all}. In this profile, we observe clear 
absorption up to +10 km s$^{-1}$ but not beyond. In the next subsection, we 
will analyze the H{\sc i} emission around this velocity to set a more stringent 
limit to the systemic velocity.

\subsection{H{\sc i} emission distribution}\label{emis}

In this section we inspect the H{\sc i} data-cube looking for signatures of the 
interaction between the SNR and the ISM. On the other hand, bearing in mind 
that the SN progenitor was a high-mass star \citep{PBW96} we also look for any 
evidence of a pre-existing structure.

\begin{figure}
\includegraphics[width=8cm]{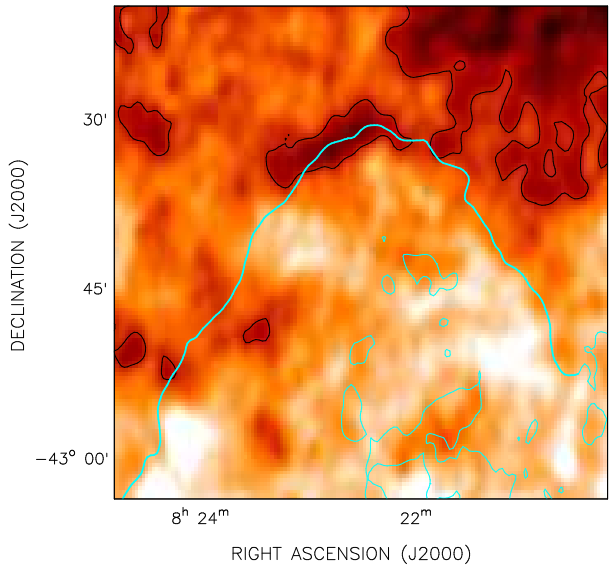}
\caption{H{\sc i} emission distribution averaged between +5.86 and +7.51 km 
s$^{-1}$. The black contour indicates the 94 K emission level, while the cyan 
contour corresponds to the 0.03 mJy beam$^{-1}$ level of the radio continuum 
emission.}
\label{hisnr}
\end{figure}

\begin{figure}
\includegraphics[width=8.5cm]{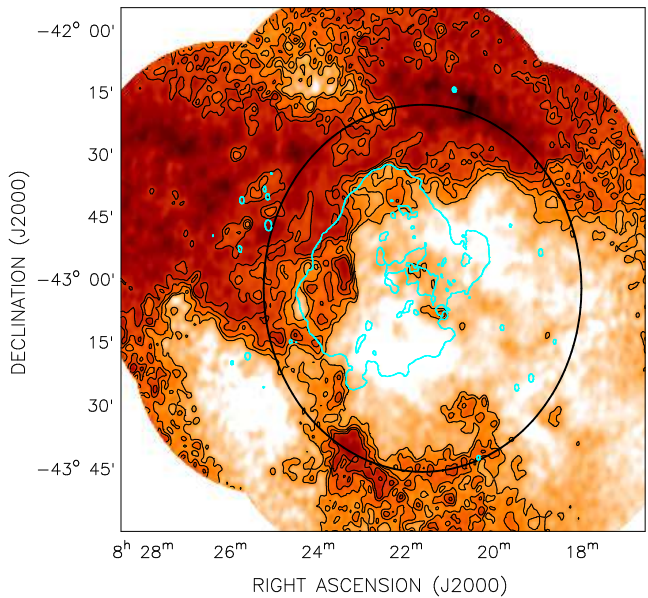}
\caption{H{\sc i} emission distribution averaged between +6.69 and +9.98 km
s$^{-1}$. Black contours are from 85 to 100 K with a step of 5 K. The cyan
contour corresponds to the 0.03 mJy beam$^{-1}$ level of the radio continuum 
emission. The ellipse shows the outline of the suggested H{\sc i} bubble.}
\label{bi}
\end{figure}

Figure \ref{sethi} shows the H{\sc i} emission distribution in the area of 
Puppis A in the velocity range from +2.56 to +29.77 km s$^{-1}$. Each box shows 
the averaged emission in three consecutive channel images, giving an effective 
velocity resolution of 2.48 km s$^{-1}$. The central velocity of each image is 
shown in the upper left corner of each box. For the sake of presentation, a 
value corresponding to the averaged brightness temperature at each radial 
velocity was first subtracted. The subtracted value is indicated
at the top right corner of each frame.

An inspection of Fig. \ref{sethi} clearly shows that the brightest regions of 
the SNR are seen in absorption in the H{\sc i} emission distribution at +2.56, 
+5.04, and +7.51 km s$^{-1}$. At v=+12.46 km s$^{-1}$, a tight correlation 
between the northern edge of the shell and the outer contour of an H{\sc i} 
concentration, could be revealing an interaction between the SNR shock front 
and the ISM at this velocity.  At the same position but at a velocity of +7.51 
km s$^{-1}$, a striking feature is observed. A close-up view of this region 
integrated between +5.86 and +8.34 km s$^{-1}$ is shown in Fig. \ref{hisnr}, 
where the excellent spatial correlation observed suggests a link between this
H{\sc i} filament and Puppis A.  From Fig. \ref{sethi} it can be said that no 
obvious H{\sc i} emission seems to be related to the SNR at velocities higher 
than v=+12.46 km s$^{-1}$.

On the other hand, since the progenitor star of Puppis A was suggested to have 
a stellar mass between 15 and 25 M${_\odot}$ \citep{hwang+08}, as mentioned 
before, we look for an H{\sc i} structure possibly created by such a star. It 
is well known that during their main sequence phase, O-type stars create, 
through their high rate of energetic photons and strong stellar winds, 
interstellar bubbles \citep[IB;][]{wea77}. These structures are detected in the
H{\sc i} emission distribution as areas of low emissivity surrounded, completely
or partially,  by  expanding shells. A structure of this kind is observed at 
around +7.5 and +10 km s$^{-1}$ (see Fig. \ref{sethi}). Figure \ref{bi} shows 
an image of the H{\sc i} emission averaged between +6.69 and +9.98  km s$^{-1}$,
where a shell-like structure is clearly observed at the location indicated by 
the ellipse. The structure seems to be open towards higher Galatic latitudes, 
probably as a consequence of the lower density found at this distance from the 
Galactic plane. The possible relation between this shell-like structure and the 
progenitor of Puppis A will be analyzed in detail in Section \ref{bi-ss}.

\section{Discussion}\label{Disc}

As can be seen in Fig.  \ref{sethi}, the H{\sc i} distribution is far from
uniform over Puppis A. This is also reflected in the difference observed in
the profiles towards the compact sources 1 to 5. From Fig. \ref{perfs-cs},
sources 1, 3, and 5 are clearly extragalactic since all of them show absorption 
features up to v$\sim +115$ km s$^{-1}$. Surprisingly, according to the spectral
index, only source 5 could have been unmistakably asserted as extragalactic.
Source 2 also has a spectral index typical of extragalactic sources, however
the last absorption feature lies at only v=$\sim +45$ km s$^{-1}$. It is likely
that absorption features at very high velocities (around +100 km s$^{-1}$) are
masked by an overestimated error band, or the H{\sc i} emission at such 
velocities in this direction is not strong enough to produce observable 
absorption. Finally, source 4, the weakest one (see last column in Table
\ref{cat-cs}), does not show obvious absorption at velocities higher than $\sim 
+10$ km s$^{-1}$. The reason could be addressed either with the large error 
band or with a local origin for this source, which could even be associated 
with Puppis A, as suggested by \citet{pap1}. Neither the absorption profile nor 
the spectral index enable us to derive a conclusive explanation about its 
origin.

While we have shown in Fig. \ref{pup_all} that Puppis A shows absorption only
up to v$\sim +10$ km s$^{-1}$, the bright eastern background source projected 
on the shell (labeled G260.72-3.16 in this paper) shows absorption at v$\sim$ 
+40, +60, and +115 km s$^{-1}$, and marginally at v$\sim$ +12.5, +22.5, and 
+105 km s$^{-1}$ (Fig. \ref{pup_east}). We note that, except for source 4, all 
other extragalactic 
sources show clear absorption around v=+12 km s$^{-1}$, with a confidence much 
better than 3$\sigma$. The abscence of an analogous feature in the absorption 
profile towards the whole remnant, even though the emission peak at this 
velocity is comparable to that at +8 and +10 km s$^{-1}$, implies that v$\sim 
+12$ km s$^{-1}$ can be reasonably set as an upper limit to the systemic 
velocity of Puppis A. The tight correlation between the northern edge of the 
shell and the H{\sc i} at v=+12.46 km s$^{-1}$ mentioned in Sect. \ref{emis} 
could be revealing an interaction between the SNR shock front and the ISM at 
this velocity. Considering this possibility, we will adopt a conservative 
estimate of the systemic velocity of v$= +10.0 \pm 2.5$ km s$^{-1}$. This range 
encloses also the bubble found at v$\sim +8$ km s$^{-1}$ mentioned in Sect. 
\ref{emis}, which will be analyzed later. In this scenario, the H{\sc i} 
filament at the northern edge of the shell (Fig. \ref{hisnr}) might correspond 
to material pushed forward and compressed by the shock front.

\subsection{Consequences of the new systemic velocity}\label{newdist}

Previous H{\sc i} and molecular studies \citep{gd+ma88,EMR+95,EMR+03} suggested
a systemic velocity of around +16 km s$^{-1}$ for Puppis A. At this velocity, 
we observe an intensity gradient towards the Galactic plane, whose tilt is
indicated through the $b = -3^{\circ}$ line (left upper box in Fig. 
\ref{sethi}). Hence, it appears that the H{\sc i} emission here represents the 
gas density distribution of the Galactic plane. Our new data show that at v=+16 
km s$^{-1}$, a good spatial correlation, although not in detail, is apparent 
{\it only} for the eastern border of the SNR, which runs parallel to the 
Galactic plane. A similar morphological match also appears up to $\sim +20$ km 
s$^{-1}$. However, there is no evidence to help disentangle whether this 
apparent correlation is due to a physical interaction between Puppis A and the 
ISM or just a projection effect.

In the light of the new absorption results, it is clear that the lower limit 
for the systemic velocity of Puppis A is $\sim$ +10 km s$^{-1}$. As discussed 
above, a comparison with absorption spectra towards other field sources seems 
to set an upper limit at v$\sim +12.5$ km s$^{-1}$. Hence, although the +16 km 
s$^{-1}$ systemic velocity cannot be ruled out, we will follow the hypothesis 
of the alternative velocity v$= +10 \pm 2.5$ km s$^{-1}$.

The new systemic velocity proposed in this paper has implications in the 
explanation of the morphology of Puppis A. Between v=+7.51 and +12.46 km 
s$^{-1}$ (Fig. \ref{sethi}), there are no traces of any thick wall of material 
capable of explaining the flattened Eastern edge of the SNR shell. The reason 
for such a morphology should be sought elsewhere, like the local ISM magnetic 
field. In this regard, we recall that the flattening is parallel to the 
Galactic Plane (Fig. \ref{sethi}), and \citet{EMR+13} have shown that in the 
case of SN 1006 the ISM magnetic field is tilted in such a direction. 

A major consequence of adopting a lower systemic velocity for Puppis A is the 
change in the kinematic distance. At v$\, \simeq +16$ km s$^{-1}$, the 
corresponding distance would be 2.2 kpc acording to the Galactic rotation model 
of \citet{FBS89}. This value has been widely used in the literature. At the new 
systemic velocity suggested in this paper, $+10.0 \pm 2.5$ km s$^{-1}$, the 
distance is re-determined as $d = 1.3 \pm 0.3$ kpc, where the quoted uncertainty
is based solely on applying the model to the velocity range assumed. An 
alternative Galactic rotation model proposed by \citet{BB93} produces a rather 
different result: $d = 2.0 \pm 0.3$ kpc which, incidentally, is very close to 
the distance thoroughly used so far. This model has the advantage that all 
longitudes are covered, while  the dataset used by \citet{FBS89} presents a gap 
from 245$^\circ$ to 351$^\circ$. However, \citet{BB93} include three sources 
approximately describing a $\sim 1$ degree arc around Puppis A: BBW129, BBW141A,
and BBW149, which have very close velocities (v$_{\rm LSR} = +9.4,\, +10.0$ and 
+8.1) but very discrepant distances: 0.59, 1.85, and 2.49 kpc respectively. 
Clearly, the dispersion in distances for such close velocities indicates that 
strong non-circular motions are taking place in this region of the Galaxy, as 
remarked by the authors. 

As an independent estimate of the distance, we used the model developed by 
\citet{chen+99} which associates the colour excess $E(B-V)$ of a certain source
with its distance \citep[see e.g.][]{EMR+06+}. We assume the standard relation
between the optical extinction $A_{\rm v}$ and the reddening 
\citep[e.g.][]{mat90}
\begin{equation}\hskip 3 cm
E(B-V) = {A_{\rm v}\over 3.1}
\label{ebv}
\end{equation}
for diffuse dust, and apply the $A_{\rm v} - N_{\rm H}$ relation derived by 
\citet{go09} for SNRs:
\begin{equation}\hskip 2 cm
N_{\rm H} = (2.21 \pm 0.09) \times 10^{21} A_{\rm v},
\label{nHAv}
\end{equation}
where $N_{\rm H}$ is the hydrogen column density in cm$^{-2}$ based on 
measurements of X-ray extinction. \citet{gh09} analyzed Newton X-Ray 
Multi-Mirror Mission observations towards the CCO in Puppis A and found $N_{\rm 
H} = (4.80 \pm 0.06) \times 10^{21}$ cm$^{-2}$, hence $E(B-V) = 0.70 \pm 0.04$. 
To apply the reddening-distance model, we estimate the total reddening 
in the direction of Puppis A produced by the Galactic Plane along the line of 
sight to be $1.58 \pm 0.05$ mag \citep{sf11}. The other two parameters used by 
the model are the distance of the Sun to the Galactic Plane, $z_\odot$, and the 
scale height of the Galactic Plane absorbing dust, $h$. Replacing both 
parameters by $h = 117.7 \pm 4.7$ pc \citep{Kos+14} and $z_\odot = 19.6 \pm 
2.1$ pc \citep{Reed06} respectively, the distance to Puppis A turns out to be 
$d = 1.2 \pm 0.2$ kpc, in excellent agreement with the value obtained through 
the Galactic rotation model of \citet{FBS89}, which we will adopt hereafter 
since it contains this latter result as well. A 1-kpc distance had also been 
derived by \citet{Z+78} based on the absorption column density of cold gas 
which yielded the best fit to the continuum X-ray spectrum. 

The closer distance would have significant implications in the interpretation 
of the CCO in Puppis A. \citet{Becker+2012} computed its proper motion to be
$71 \pm 12$ mas yr$^{-1}$ based on 4 different measurements over one decade. 
At a distance of $1.3 \pm 0.3$ kpc, this velocity translates into $440 \pm
175$ km s$^{-1}$. This value is in excellent agreement with the mean pulsar 
space velocity of $440 \pm 40$ km s$^{-1}$ \citep{hllk05} and, since it does 
not exceed 500 km s$^{-1}$, there would be no need to invoke a hydrodynamic 
recoil mechanism, as proposed by \citet{Becker+2012} to explain the anomalously 
high velocity kick. 

With the distance reduced to 1.3 kpc, the radius of Puppis A would be 
$\lesssim 10$ pc, and the expansion velocity, $v_{exp} = \frac{2}{5} R/t 
\simeq 750$ km s$^{-1}$, where the $\frac{2}{5}$ factor applies if the SNR is 
assumed to be in the Sedov phase. These values for size and expansion velocity, 
which would be 70\% higher with the previous distance of 2.2 kpc, are more in 
line with typical sizes and velocities of SNRs older that a thousand years 
\citep{chiad+15}. Overall, although the $\sim 2$ kpc distance cannot be 
completely ruled out, there are plenty of arguments in favor of Puppis A
having a distance closer to 1 kpc.

\subsection{An interstellar bubble created by the progenitor of Puppis A? }\label{bi-ss}

\begin{table*}
\caption{Stellar parameters}\label{stellar}
\vskip 0.25truecm
\centering
\begin{tabular}{c c c c c c}
\hline\hline
Spectral Type & Mass loss rate   & Wind Velocity & Main-sequence lifetime & Radius achieved & Distance $^a$\\
 & $\times 10^{-8}$ M$_{\odot}$\, yr$^{-1}$ & km\, s$^{-1}$ & Myr & pc & kpc\\
\hline
O7 & 1.9 & 2412 & 6.3 & 14.6 & 1.2\\
O7.5 & 1.4 & 2388 & 7.3 & 14.5 & 1.2 \\
O8 & 1.0 & 2364 & 8.0 & 14.0 & 1.1 \\
O8.5 & 0.7 & 2355 & 8.2 & 12.9 & 1.1 \\
O9 & 0.5 & 2332 & 9.0 & 12.5 & 1.0 \\
O9.5 & 0.4 & 2311 & 10.5 & 12.4 & 1.0 \\

\hline
\end{tabular}
\begin{list}{}{}
 \item {$^a$} Distance necessary to reproduce the linear radii in column 5 for
an angular radius of 0\fdg6.
\end{list}
\end{table*}

\begin{figure}
\includegraphics[width=11cm]{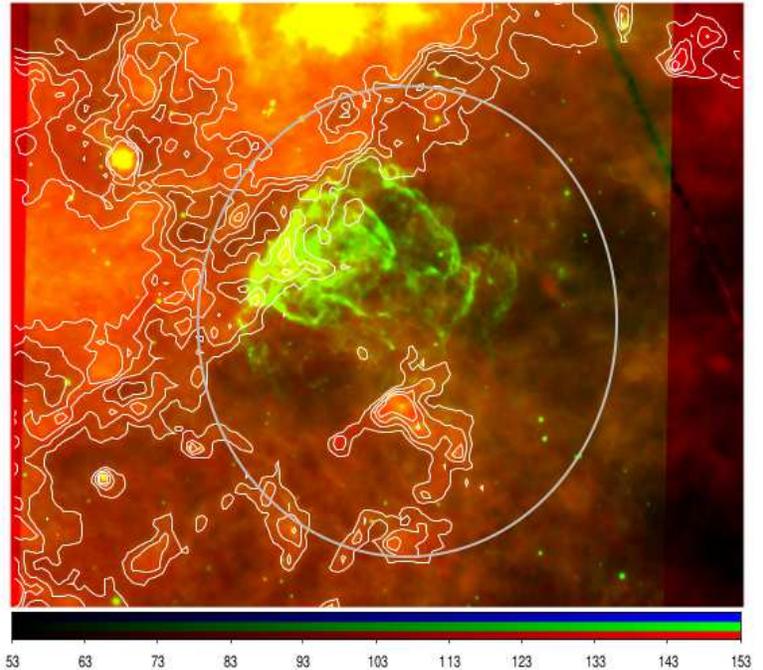}
\caption{Composite image of the region of Puppis A. The image shows WISE 22 
$\mu$m emission in green and the AKARI 140 $\mu$m in red and contours, which 
correspond to the emission at 120, 130, 140, and 150 MJy sr$^{-1}$. The ellipse 
is the same as Fig. \ref{bi} and shows the outline of the suggested H{\sc i} 
bubble.}
\label{ir}
\end{figure}

As shown in Section \ref{emis}, a shell-like structure is observed in the 
velocity range from about +6.7 to +10 km s$^{-1}$. From Fig. \ref{bi}, we can 
estimate for the shell an angular radius of about 0\fdg6. In what follows we 
will discuss if this structure could have been created by the progenitor of 
Puppis A. In this regard, it is important to note that the structure is 
observed in a narrow velocity range, less than 5 km s$^{-1}$, implying a low 
expansion velocity, which is expected if the shell was blown by a star that has 
ceased its mass ejection activity long ago, after exploding as a SN.

As pointed out by \citet{hwang+08}, the progenitor had a mass between 15 and 
25 M$_{\odot}$, which implies a spectral type between O7 and O9.5 during the 
main sequence phase \citep{mar05}. Taking this range into account we now 
analyze whether such a star could have created the observed H{\sc i} shell. To 
this end we need to assume a value for the original ambient density, i.e. the 
density found by the high-mass star during its main-sequence phase. As a rough 
estimate, we use the Galactic density distribution proposed by \citet{kal08} 
for Galactocentric distances in the range $7 \le R_{GC} \le 35$ kpc, given by
\begin{equation}\hskip 2.5 cm
n  \sim n_o\, e^{-(R_{GC} - R_{\odot})/ R_n}
\label{amb_dens}
\end{equation} 
where $n_o = 0.9$ cm$^{-3}$, $R_{\odot} = 8.5$ kpc, and $R_n$ = 3.15 
kpc is a scale length. Adopting $R_{GC}$ = 9 kpc for Puppis A, we obtain $ n 
\sim 0.8$ cm$^{-3}$.  

The size achieved for a structure created by the wind energy injected by a 
given star depends mainly on the wind properties. The mass loss rates 
($\dot{M}$) and terminal wind velocities ($v_w$) assumed in this paper for the 
different possible spectral types are given in Table \ref{stellar}, and were 
estimated from \citet{dej88},
\begin{equation}
\log \dot{M} = 1.769\, \log \frac{L}{L_{\odot}} - 1.676 \, \log T_{\rm eff} - 8.158
\label{Mdot}
\end{equation} 

\begin{equation}
\log v_w = 1.23 - 0.3 \log \frac{L}{L_{\odot}} + 0.55 \log \frac{M}{M_{\odot}} 
+ 0.64 \log T_{\rm eff}
\label{Vwind}
\end{equation} 
where $L$ is the stellar luminosity expressed in solar luminosity units, $M$ is 
the stellar mass expressed in solar masses,  and $T_{\rm eff}$ is the effective 
stellar temperature in Kelvin. For each spectral type, the basic stellar 
parameters ($L$, $M$, and $T_{\rm eff}$) were assumed to be the values given by 
\citet{mar05}.

The strong impact of the star upon the ISM takes place during the whole 
main-sequence phase, whose duration depends on the stellar mass. The adopted 
lifetimes were taken from the tables of \citet{sch92} and are shown in column 4
of Table \ref{stellar}. To estimate the radius reached by  the structure at the 
end of the main-sequence phase we follow the evolution model of \citet{wea77} 
which, for the constant wind momentum phase, gives
\begin{equation}\hskip 2.5 cm
R = 0.83 \, (\frac{\dot{M} \, V_w}{\rho_0})^{1/4} \, t^{1/2}
\label{radio}
\end{equation} 
where $\rho_0$ is the density of the ambient medium assumed uniform. The values 
obtained are given in the 5th column of Table \ref{stellar}.

As mentioned above, the observed structure has an angular radius of about 
0\fdg6. In the last column of Table \ref{stellar}, the distance needed for the 
linear radius of the shell to be  equal to the radius estimated through Eq.
\ref{radio} (column 5 in Table \ref{stellar}) are given. As can be seen, the
derived distance is from 1.0 to 1.2 kpc, in excellent agreement with the new 
estimated distance for Puppis A (see Section \ref{newdist}). Thus, in view of 
these results it seems completely possible that the star progenitor of Puppis A 
has created the observed H{\sc i} bubble.

Finally, we inspected the infrared emission in the area to see whether the 
observed shell has an IR counterpart. Figure \ref{ir} shows a two colour 
image of the area. The Wide-field Infrared Survey Explorer 
\citep[WISE,][]{wri10} 22 $\mu$m emission is shown in green and the AKARI 
\citep{tak15} 140 $\mu$m, in red. From this image it is clear  that the 
infrared emission related to Puppis A is higher at 22 $\mu$m, while at longer 
wavelengths, where the IR emission traces cooler dust components, an 
arc-shaped structure is detected showing a good morphological correlation 
with the southern part of observed HI bubble (indicated by the ellipse).
A striking infrared structure is detected inside the ellipse at about ($\alpha, 
\delta$) = ($8^{\rm h} 21^{\rm m} 45^{\rm s}$, $-43^{\circ} 18^{\prime}$) (see 
Fig. \ref{ir}). At this location, there is a dark cloud catalogued as 
DCld\,260.6-3.7 by \citet{otr00}. However, since this source has velocity 
components at  $v = +4.5 $ and $+6.4$ km s$^{-1}$, it is clearly a foreground
cloud not related to Puppis A.

\section{Conclusions}

We have performed H{\sc i} observations using the ATCA and subsequently 
combined these data with single-dish observations, to produce an H{\sc i} 
mosaic around the SNR Puppis A. The resultant data has high sensitivity (2-9
mJy\,beam$^{-1}$), good spatial resolution ($118\farcs3 \times 88\farcs9$) and 
a spectral resolution of 0.82 km\,s$^{-1}$. We use these data to investigate 
the physical properties of Puppis A in two ways:
\begin{enumerate}
\item We focus on five unresolved continuum point sources from \citet{pap1}
and investigate the H{\sc i} absorption against the background continuum of 
these sources. In sources 1, 3, and 5, we find H{\sc i} absorption at velocities
up to $\gtrsim$100\,km\,s$^{-1}$, as well as negative spectral indices 
confirming their extragalactic origin. In addition to this, we also identify 
G260.72-3.16, which is found on the eastern edge of the SNR, as a background 
source, probably extragalactic, approximately 2 arcminutes long and oriented 
north-south. We have filtered out the contribution from these background 
sources and discuss that the best match for the velocity of the SNR is 
$+10.0\pm2.5$\,km\,s$^{-1}$, which is significantly smaller than found by 
previous work.
\item We investigated the morphology of H{\sc i} emission over Puppis A and 
surrounding regions. We find a good match of continuum and H{\sc i} emission 
morphologies at velocities of +7.51, +9.98, and +12.46\,km\,s$^{-1}$. This 
provides further evidence that the systemic velocity of Puppis A is found 
within this range although a moderate match at higher velocities does not 
permit to fully rule out the previously accepted velocity of +16 km
s$^{-1}$. We also see some evidence for a shell-like structure in the 
H{\sc i} emission in the velocity range of +6.69 to +9.98\,km\,s$^{-1}$. This 
shell structure appears to surround the SNR, seen in radio continuum emission. 
We interpret this shell as a bubble in the H{\sc i} emission that was created 
by the supernova explosion.
\end{enumerate}

Based on our alternative systemic velocity of Puppis A of $+10.0\pm2.5$\,km 
s$^{-1}$, we compare this velocity to Galactic rotation curves in order to 
determine a kinematic distance. We therefore revise the estimated distance of 
Puppis A to $1.3\pm0.3$\,kpc. This distance is confirmed by a comparison to the 
distance separately estimated based on a colour excess model.

Given the revised distance, we calculate a proper motion velocity of $440 \pm
175$\,km\,s$^{-1}$ for the CCO, which is significantly smaller than previous 
estimates and does not require a hydrodynamic recoil mechanism to accelerate 
it to unusually high velocities. We estimate the radius of Puppis A to be 
$\sim$10\,pc, with an expansion velocity of $\sim$750\,km\,s$^{-1}$, 
compatible with a SNR older than 1000 years.

\section*{Acknowledgements}

We acknowledge Peter Kalberla for providing us with Parkes H{\sc i} data from
the Galactic All-Sky Survey (GASS) and fixing calibration problems in its
original version. We appreciate useful discussions with Laura Richter, Timothy 
Shimwell and Mark Wieringa for solving the mosaic field name length
bug, and Jamie Stevens for support to solve the 1 MHz zoom mode bug. This 
research was partially funded by CONICET grants PIP 114-200801-00428 and 
112-201207-00226. The Australia Telescope Compact Array is part of the 
Australia Telescope National Facility which is funded by the Commonwealth of 
Australia for operation as a National Facility managed by CSIRO. EMR and SC
are members of the Carrera del Investigador Cient\'\i fico of CONICET, 
Argentina.

\bibliographystyle{mn2e}
\bibliography{apj-jour,pupHI}{}

\bsp

\label{lastpage}

\end{document}